\documentclass[12pt]{article}  
\usepackage{cite}
\usepackage{epsfig}
\usepackage{graphicx}
\usepackage{amsmath}
\usepackage{amssymb}
\usepackage{ulem}
\usepackage{mdwlist} 
\usepackage{color}  
\usepackage[OT2, T1]{fontenc}
\usepackage[russian,english]{babel}

\usepackage{a41}
\usepackage{color} 
\usepackage[rflt]{floatflt}
\usepackage{float}
\usepackage{slashed}

\usepackage{array}

\usepackage[english]{babel}

\usepackage{url}

\usepackage{amsmath, amsthm, amssymb}
\newtheorem{thm}{Theorem}[section]

\newtheorem{definition}[thm]{Definition}

\newcommand{\Li}{{\rm Li}}
\newcommand{\HA}{{\rm H}}

\newcommand{\Mvec}{{\rm\bf M}}

\newcommand{\SigmaP}{\texttt{Sigma}}

\newcommand{\ep}{\varepsilon}

\usepackage{rotating}
\newcommand{\shuffle}{\, \raisebox{1.2ex}[0mm][0mm]{\rotatebox{270}{$\exists$}} \,}
\usepackage{graphicx}

\newcounter{mmacnt}
\def\restartmma{\setcounter{mmacnt}{0}}
\restartmma \catcode`|=\active
\def|#1|{\mathrm{#1}}
\catcode`|=12
\newenvironment{mma}{
 \par\smallskip
 \catcode`|=\active
 \parskip=0pt\parindent=0pt 
 \small
 \def\In##1\\{%
\def\linebreak{\hfill\break\null\qquad}%
\refstepcounter{mmacnt}
\hangindent=2.5em\hangafter=0
\leavevmode
\llap{\tiny\sffamily n[\arabic{mmacnt}]:=\kern.5em}%
\mathversion{bold}\footnotesize$\displaystyle##1$\normalsize
\mathversion{normal}\par
 }%
 \def\Print##1\\{%
\def\linebreak{\hfill\break}%
\hangindent=2.5em\hangafter=0
\leavevmode ##1\par}%
 \def\Out##1\\{%
\def\linebreak{$\hfill\break\null\hfill$}%
\kern\abovedisplayskip\par
\hangindent=2.5em\hangafter=0
\leavevmode
\llap{\tiny\sffamily Out[\arabic{mmacnt}]=\kern.5em}
\footnotesize$\displaystyle##1$\normalsize\hfill\null\par
\kern\belowdisplayskip
 }%
 \def\Warning##1##2\\{%
\def\linebreak{\hfill\break}%
\hangindent=2.5em\hangafter=0
\leavevmode
{\scriptsize##1 : ##2}\par}%
}{%
 \par\smallskip
}

\usepackage{color}

\newenvironment{fshaded}{%
\MakeFramed {\FrameRestore}
}%
{\endMakeFramed}

\allowdisplaybreaks[4]

\begin{document}
\setlength{\baselineskip}{0.515cm}
\sloppy
\thispagestyle{empty}
\begin{flushleft}
DESY 18--045
\hfill 
\\
DO--TH 18/06 \hfill Int.J.Mod.Phys. A33 (2018) no.17, 1830015 \\
April 2018\\
\end{flushleft}

\mbox{}
\vspace*{\fill}
\begin{center}

{\LARGE\bf Analytic Computing Methods for Precision} 

\vspace*{3mm} 
{\LARGE\bf Calculations in Quantum Field Theory}

\vspace{3cm}
\large
Johannes~Bl\"umlein$^a$ and 
Carsten~Schneider$^b$

\vspace{1.cm}
\normalsize
{\it  $^a$ Deutsches Elektronen--Synchrotron, DESY,}\\
{\it  Platanenallee 6, D-15738 Zeuthen, Germany}
\\

\vspace*{3mm}
{\it $^b$~Research Institute for Symbolic Computation (RISC),\\
                          Johannes Kepler University, Altenbergerstra\ss{}e 69,
                          A--4040, Linz, Austria}\\


\end{center}
\normalsize
\vspace{\fill}
\begin{abstract}
\noindent
An overview is presented on the current status of main mathematical computation methods for the multi-loop 
corrections to single scale observables in quantum field theory and the associated mathematical number and 
function spaces and algebras. At present massless single scale quantities can be calculated analytically in 
QCD to 4-loop order and single mass and double mass quantities to 3-loop order, while zero scale quantities 
have been calculated to 5-loop order. The precision requirements of the planned measurements, particularly 
at the FCC-ee, form important challenges to theory, and will need important extensions of the presently 
known methods.   
\end{abstract}

\vspace*{\fill}
\noindent

\newpage

\vspace*{1mm}
\noindent
\vspace*{-2cm}
\section{Introduction}
\label{BS:sec:1}

\vspace*{1mm}
\noindent
Precision calculations of radiative corrections in the Standard Model play an important 
role in determining its fundamental parameters, like the particle masses, coupling constants 
and mixing parameters and are important to reveal possible deviations due to new physics. These 
calculations
are very challenging and have to be more precise than the experimental accuracy to be reached.
At present the necessary calculations, both in the case of the physics at HERA and the LHC, for typical 
quantities, last about one to {two decades} until the final results are obtained. A similar and 
probably larger period is estimated to be necessary to meet the requirements at a future circular 
collider, such as the FCC \cite{FCC}. This includes both the necessary development of new technologies, 
and to a lesser extent, also substantial computation times.

In this survey we will summarize the status reached in the field of analytic radiative corrections mostly for 
inclusive 
single and double scale quantities, widely concentrating on QED and QCD, and discuss the different
techniques available at the higher loop level at present. Currently we are at the beginning of a 
discussion of what
is needed at the theoretical side to meet the planned experimental accuracy for key processes, e.g. at the 
FCC-ee. This r\'esum\'e of computational methods is meant to be one asset to explore in which way these 
technologies have to develop in the future, concerning the computational complexity and the different 
theoretical, mathematical and computer algebraic methods, as far as this is currently foreseeable. 

The mathematical representation of the one-loop case has been given by 't Hooft and Veltman in \cite{tHooft:1978jhc,
tHooft:1973wag,Veltman:1994wz} during the 1970ies and the analytic representation of 5- and higher point 
functions were given in  \cite{Denner:2002ii,Binoth:2008uq,Fleischer:2010sq,Fleischer:2011hc} and 
references therein.
Also the one-loop off-shell amplitudes are known in full detail \cite{Fleischer:2003rm,Bluemlein:2017rbi,
Blumlein:2017a}. At general space-time dimensions these quantities are represented in terms of higher
hypergeometric functions, such as the hypergeometric function itself \cite{HYPKLEIN,HYPBAILEY,SLATER1},  
Appell functions \cite{APPEL1,APPEL2,KAMPE1,EXTON1,EXTON2,SCHLOSSER,SRIKARL}
and the Lauricella-Saran functions \cite{Lauricella:1893,Saran:1954,Saran:1955,SRIKARL}.
For the multi-leg corrections, containing more scales, mostly numerical methods have to be used at present, cf. 
\cite{Blumlein:2012gu} for a survey.

The most far reaching result on radiative corrections at LEP have been the complete $O(\alpha^2)$ 
QED corrections
around the $Z$ resonance \cite{Berends:1987ab}, cf. also~\cite{Blumlein:2011mi}. During the 1990ies most 
of 
the single and some of the double scale two-loop corrections for different processes have been calculated in 
QCD, cf. e.g.~\cite{Hamberg:1990np,Zijlstra:1992qd,Buza:1995ie}. At the mathematical side, polylogarithms 
\cite{LEWIN1,LEWIN2} and Nielsen integrals \cite{NIELSEN,KMR70,Kolbig:1983qt,Devoto:1983tc}
were used, leading to complicated argument structures. 

Around 1998\footnote{For computation techniques until 1998, see 
Ref.~\cite{Harlander:1998dq}.} a more systematic account was needed to be able to perform more advanced 
calculations and to simplify
what has been reached at the 2-loop level. This led to the introduction of the nested harmonic sums
\cite{Vermaseren:1998uu,Blumlein:1998if} and the harmonic polylogarithms \cite{Remiddi:1999ew}. Within this 
framework various important massless \cite{Moch:2004pa,Vogt:2004mw,Vermaseren:2005qc,Ablinger:2017tan}
and massive \cite{Bierenbaum:2008yu,Ablinger:2010ty,Ablinger:2014vwa} 3-loop corrections could be calculated.
In later computations extensions of these function spaces to generalized harmonic polylogarithms 
\cite{Moch:2001zr,Ablinger:2013cf}, cyclotomic harmonic sums and polylogarithms \cite{Ablinger:2011te}, and 
root-valued iterated integrals and finite weighted (inverse) binomial sums \cite{Ablinger:2014bra} were 
necessary. Finally, also elliptic integrals \cite{TRICOMI} appear in massive Feynman integral calculations 
\cite{Broadhurst:1993mw,
Laporta:2004rb,Bloch:2013tra,Adams:2013kgc,Adams:2014vja,Adams:2015gva,Adams:2015ydq,
Adams:2016xah,Ablinger:2017bjx}, and modular forms and functions, see e.g.~\cite{SERRE,COHST,ONO1}, play a 
role
in describing the corresponding physical results. In this paper we will mainly discuss zero-, one-, and 
a few two-scale processes. In general, processes with more than two scales are important as well. However, the
systematic mathematical study of the corresponding analytic master integrals, beyond the one-loop case, 
is at the 
beginning only.

The paper is organized as follows. In Section~\ref{BS:sec:2} we give an outline of a systematic way to classify 
Feynman integrals. It will be given studying the structure of the associated system of differential equations 
w.r.t.\ its degree of factorization. General computation methods are summarized in Section~\ref{BS:sec:2A}. In 
Section~\ref{BS:sec:3} we describe a series of analytic integration methods for Feynman diagrams. There are 
universal corrections to scattering processes, characterized by logarithmic enhanced terms, which we 
describe  in Section~\ref{BS:sec:F} using the structure function method. It relies on the 
renormalization group equation for factorization. The different kind of loop corrections can be presented by 
a series of function and special number spaces which will be reviewed in Section~\ref{BS:sec:4}. Here we will
mainly concentrate on the spaces for first-order factorizing systems. 
In Section~\ref{BS:sec:5} we will discuss non-first order factorizing systems and present an outlook in 
Section~\ref{BS:sec:6}.
\section{A Systematic Way to Classify Feynman Integrals}
\label{BS:sec:2}

\vspace*{1mm}
\noindent
Feynman parameter integrals can be calculated directly by applying successive binomial
\begin{eqnarray}
\label{eq:BIN}
(A(x_i) + B(x_i))^k = \sum_{l=0}^k \binom{k}{l} A^l(x_i) B^{k-l}(x_i),~k \in \mathbb{N},
\end{eqnarray}
and Mellin-Barnes  \cite{BARNES1,MELLIN1}
\begin{eqnarray}
\label{eq:MB}
\frac{1}{((A(x_i) + B(x_i))^\lambda} = \frac{1}{2\pi i} \int_{-i \infty}^{+i \infty} dz 
\frac{B(x_i)^z}{A(x_i)^{\lambda + z}} \frac{\Gamma(\lambda+z) \Gamma(-z)}{\Gamma(\lambda)},
~\lambda \in \mathbb{R}, \lambda > 0, 
\end{eqnarray}
decompositions of the integrand implemented in different packages 
\cite{Czakon:2005rk,Smirnov:2009up,Gluza:2007rt}.
Here $\{x_i\}$ denotes the set of Feynman parameters and $A$ and $B$ are polynomials. These 
representations allow one to perform all Feynman parameter integrals. The remaining Mellin-Barnes integrals
(\ref{eq:MB}) have to be carried out using the residue theorem.
In this way one obtains multi-sum representations, as has been outlined in 
Ref.~\cite{Blumlein:2010zv}. However, the remaining summation problems may turn out to be very difficult
to be solved using difference field and ring theory \cite{Karr:1981,Bron:00,Schneider:01,Schneider:04a,Schneider:05a,Schneider:05b,
Schneider:07d,Schneider:10b,Schneider:10c,Schneider:15a,Schneider:08c,Schneider:08d,Schneider:08e}, algorithmically 
implemented in the package {\tt Sigma} \cite{Schneider:2007a,Schneider:2013a}. It may even turn out 
that the corresponding nested sums cannot be solved using difference field and ring theory in terms of indefinite nested sums defined over hypergeometric  
products because the associated difference equation does not factorize sufficiently. 

It thus seems that there are apparently no general methods known {\it a priori}, to evaluate a given 
general Feynman diagram 
analytically. Therefore we want to attempt at least to find general classes first, for which the corresponding 
mathematical solution methods can be found or be developed, along with classifying the different degrees of complexity.

One way to proceed is the following. In most of the present and future Feynman diagram calculations one 
has to use the integration-by-part (IBP) method \cite{Chetyrkin:1981qh} to reduce the problem to master 
integrals. There are various implementations
of Laporta's algorithm \cite{Laporta:2001dd}, as e.g. \cite{Studerus:2009ye,vonManteuffel:2012np,MARSEID} and 
others. Given such a system, the following definition will be convenient for further considerations.
An $n \times n$ system is said to decouple to $m$th order, $m \leq n$, if the $n$ linear differential operators that contain the solutions of the corresponding unknown functions of the homogeneous system factorize to irreducible factors where the maximal order among these factors is $m$.  
Note that decoupling algorithms such as the cyclic-vector method provide such scalar equations for the unknown 
functions~\cite{ORESYS,BCP13}. In our considerations below we will utilize Z\"urcher's algorithm~\cite{Zuercher:94} which 
might provide a refined version with several scalar equations. 
However, if the scalar equations found by any of the available decoupling algorithms factorize linearly, the system decouples to first-order, i.e., $m=1$. In particular, the system is completely solvable in terms of indefinite nested integrals (resp.\ indefinite nested sums in the recurrence case). This special case will be considered in detail in Section~\ref{BS:sec:3}. 

In this way, one can determine the degree of uncoupling of the system and
a systematic 
characteristics of sets of Feynman 
integrals can be given in terms of their master integrals. Here, the decoupling has been carried out in 
$x$-space, with $x$ being the variable of the differential equations. One may now consider a map to 
$N$-space either by a formal power series Ansatz or a Mellin transform and study the problem in terms of the
corresponding system of difference equations. Usually the uncoupling can be different here. There are cases 
know in which, e.g., 1st order decoupling can be obtained in Mellin space \cite{Ablinger:2017bjx} but not 
in $x$ 
space and vice 
versa \cite{Czakon:2008ii}. If possible, one should always consider both cases. First order factorizing 
systems can be always solved in terms of iterative integrals or in $N$-space by indefinitely nested sums.

2nd order decoupling systems will have $_2F_1$-solutions, with usually more than three singularities,
cf. e.g.~\cite{Laporta:2004rb,Ablinger:2017bjx}. Much less is known about genuine 3rd and higher order 
systems, which are, however, expected to occur at higher loop order, or in the presence of more scales. 
The computational methods to solve 1st order systems, cf. Section~\ref{BS:sec:3}, \ref{BS:sec:4}, 
are different from those of the 2nd order systems, cf.~Section~\ref{BS:sec:5}. It is therefore 
expected that at larger degrees of non-decoupling quite different mathematical methods are necessary 
for adequate analytic solutions of these radiative corrections, and may not even be completely known
in the mathematical literature w.r.t.\ the function spaces covering the corresponding representation.

After a brief consideration of general computation methods for Feynman integrals, which are standard by now, 
in the next section, we turn to the case of 1st order factorization systems.
\section{General Computation Methods}
\label{BS:sec:2A}

\vspace*{1mm}
\noindent
The present higher loop integrals make it necessary to evaluate
a large number of Feynman integrals; their number will grow in future projects significantly. 
Currently there is no method known for the renormalizable quantum field theories realized in nature to start
from any more compact structure.
The generation of the Feynman diagrams, using packages like {\tt qgraf} \cite{Nogueira:1991ex}, still seems to be 
possible as well as the calculation of the corresponding color structures using {\tt Color} 
\cite{vanRitbergen:1998pn}. Furthermore, standardized algorithms to obtain Feynman parameterizations 
exist, cf. e.g. \cite{NAKANISHI,LEFSCHETZ,Bogner:2010kv}.
However, with growing complexity, to perform the Dirac- and spin-algebra will be a challenge even 
to {\tt FORM} 
\cite{Vermaseren:2000nd, Tentyukov:2004hz, Tentyukov:2007mu, Ruijl:2017dtg}. These packages need steady 
maintenance and will need further refinements. It will be indispensable to reduce the corresponding scalar 
(or tensor) integrals to master integrals. Here, at present, we have a series of efficient algorithms 
like 
{\tt Air, FIRE, Crusher, REDUZE} \cite{Anastasiou:2004vj,Smirnov:2008iw,MARSEID,Studerus:2009ye,vonManteuffel:2012np}.
These algorithms definitely will need further extensions and improvements to face future 
problems.
Before being able to embark to the analytic calculation of master integrals one has to have numerical tools
of sufficient accuracy to evaluate the master integrals setting their remaining parameters to specific values.
Here the method of sector decomposition and related numerical methods are used, see e.g.~\cite{Hepp:1966eg,
Binoth:2000ps,Nagy:2006xy,Anastasiou:2007qb,Smirnov:2008py,Carter:2010hi,Smirnov:2009pb,Becker:2010ng,
Becker:2012aq,Becker:2011vg,Smirnov:2015mct,Borowka:2018dsa}. These methods have also to be refined further to 
meet future tasks. 
\section{A Series of Analytic Integration Methods}
\label{BS:sec:3}

\vspace*{1mm}
\noindent
We turn now to the analytic integration of Feynman integrals and master integrals, respectively.
In the following, we limit the consideration mainly to zero- and single scale processes. Most 
of the methods
lead to representations of nested finite and infinite sums, which have to be solved applying summation
techniques. These are therefore a central tool in the calculation of Feynman integrals. Non of 
the methods, which we are going to discuss in the following, has a thorough superiority over an other
in case it can be applied. As experience in larger projects shows, one method is better suited for
the calculation of certain diagrams than others. The differences in run-time can sometimes 
be quite large and e.g. amount to different weeks compared to seconds. Let us now turn to a series of 
specific integration methods.
\subsection{The PSLQ method}
\label{BS:sec:31}

\vspace*{1mm}
\noindent
The PSLQ method \cite{PSLQ} is an integer algorithm with which a possible dependence between 
certain constants, expressed by floating point numbers to a certain accuracy, can be 
determined. To do this one starts with a certain accuracy of $n$ digits and tries to find a
first relation, usually given by a sequence of rational numbers, with a certain denominator size.
One then increases the accuracy and checks whether the relation is found again.
The method has no proof character, however, it is able to establish approximate relations between 
constants at a high accuracy. If a correct relation is found, one can proof it by rigorous methods later,
which, however, is usually by far more demanding. The well-know relation
\begin{eqnarray}
\label{eq:PSLQ1}
\Li_3\left(\frac{1}{2}\right) = \frac{7}{8} \zeta_3 - \frac{1}{2} \zeta_2 \ln(2) + \frac{1}{6} \ln^3(2)
\end{eqnarray}
can be found already providing 9 digits. In this case one knows, that all participating numbers in the 
r.h.s. are irrational. Usually one needs many more digits, however, 
cf.~e.g.~\cite{Borwein:1999js,Laporta:2017okg}. 

The method can be very efficiently used if one knows the set of numbers, by which the result of a 
calculation is finally spanned. In intermediary steps many more unknown constants may appear, which cancel 
in the final result. This has been very impressively demonstrated in the calculation of the 5--loop
$\beta$--function in QCD in Ref.~\cite{Luthe:2017ttg}, where for the master integrals even large amounts 
of elliptic and probably trans-elliptic constants contribute, which all cancel finally. It was expected, 
that the 5--loop $\beta$--function depends on $\zeta$-values \cite{Blumlein:2009cf} only and then also 
known by the results of Refs.~\cite{Baikov:2016tgj,Herzog:2017ohr}.

Similarly, one makes use of this also in the method of arbitrarily high moments, 
Section~\ref{BS:sec:39}. E.g. for the calculation of the massive 3--loop operator matrix element (OME) 
$A_{Qg}^{(3)}(N)$, it is known, that their moments depend on a series of $\zeta$-values only, 
\cite{Bierenbaum:2009mv}. The master 
integrals and the general $N$-result contain elliptic integrals \cite{Ablinger:2017bjx}, on the other 
hand.

The PSLQ-method provides a great heuristic algorithm to potentially simplify zero-scale quantities in 
higher order loop calculations. 
\subsection{Guessing difference equations}
\label{BS:sec:32}

\vspace*{1mm}
\noindent
Often it is easier to calculate a large series of zero-scale quantities, e.g. the Mellin moments,
\begin{eqnarray}
\label{eq:MELLIN}
G(N) = \Mvec[f(x)](N) = \int_0^1 dx x^{N-1} f(x), 
\end{eqnarray}
of a sought single-scale quantity $G(N)$ for general values of $N$.
One therefore seeks ways to find the associated difference 
equation \cite{NOERLUND} to the set of moments $\{G(2),G(4),....,G(2m)\}, m \in \mathbb{N}$
\cite{Larin:1993vu,Larin:1996wd,Retey:2000nq,Blumlein:2004xt}. The reason why
one may hope to find a difference equation is that $G(N)$ is a function obeying a recurrence. This is the case
for many objects in quantum field theory, as e.g. (massive) operator matrix elements \cite{Bierenbaum:2009mv},
but also single-scale Wilson coefficients, Ref.~\cite{Vermaseren:2005qc}.

A possible algorithm to calculate the string of moments $\{G(k)|_{k=2}^{2m}\}$ as input is described in 
Section~\ref{BS:sec:39}. One then applies the guessing algorithm \cite{GSAGE}, now also available in {\tt 
Sage} \cite{SAGE},
exploiting the fast integer algorithms available there. At a critical number $m$ a conjectured difference 
equation is obtained, the validity of which can be tested by shifting to a further number of higher moments.
This method has been applied in Ref.~\cite{Blumlein:2009tj} to obtain from more than 5000 moments the massless
unpolarized 3-loop anomalous dimensions and Wilson coefficients in deep-inelastic scattering
\cite{Vogt:2004mw,Moch:2004pa,Vermaseren:2005qc}.
Recently, the method has been applied {\it ab initio} in the calculation of 3-loop splitting functions 
\cite{Ablinger:2017tan}
and the massive 2- and 3-loop form factor \cite{Ablinger:2017tan,FORMF3}. In the case of a massive operator 
matrix element 8000 moments \cite{Ablinger:2017ptf} could be calculated and difference equations 
were derived for all 
contributing color and $\zeta$-value structures. Using the summation methods implemented in the package {\tt 
Sigma} \cite{Schneider:2007a,Schneider:2013a} one can solve these difference equations, whenever they are first 
order factorizing, or separate their first order factorizing parts, in case they are not, 
cf.~Section~\ref{BS:sec:37}. In Section~\ref{BS:sec:39} we will describe in which way the corresponding 
input-moments can be obtained.
\subsection{Generalized hypergeometric functions}
\label{BS:sec:33}

\vspace*{1mm}
\noindent
In calculating simpler Feynman integrals, or those with a partial factorization to simpler structures, the Feynman parameter
representations exhibit integrand structures of Euler Beta-functions, the hypergeometric functions
or their generalization \cite{HYPKLEIN,HYPBAILEY,SLATER1} in general space-time dimensions, e.g. $D = 4 + 
\varepsilon$. Up to the level of massless and single mass two-loop integrals, 
cf.~\cite{Hamberg:1990np,HAMBERG,Buza:1995ie,Bierenbaum:2007qe}, these representations are usually sufficient.

There is a hierarchy of $_{p+1}F_p$ functions, the first of
which read
\begin{eqnarray}
B(a_1,a_2)           &=& \int_0^1 dt~t^{a_1-1} (1-t)^{a_2-1}
\\
\label{eq:2F1}
_2F_1(a_1,a_2,b_1;x) &=& 
\frac{\Gamma(b_1)}{\Gamma(a_2)\Gamma(b_1-a_2)} \int_0^1 dt~t^{a_2-1} 
(1-t)^{b_1-a_2-1} (1-tx)^{-a_1}
\\
_3F_2(a_1,a_2,b_1;x) &=& \frac{\Gamma(b_2)}{\Gamma(a_3) \Gamma(b_2-a_3)} \int_0^1 dt~t^{a_3-1} 
(1-t)^{-a_3+b_2-1}~_2F_1(a_1,a_2,b_1;tx).
\end{eqnarray}
Here the parameters $a_i, b_i$ are such, that the corresponding integrals exists, \cite{WITWAT}.
At 3-loop order, there are topologies which can be well described by Appell functions 
\cite{APPEL1,APPEL2,KAMPE1,EXTON1,EXTON2,SCHLOSSER,Anastasiou:1999ui,Anastasiou:1999cx,SRIKARL}, 
cf.~\cite{Ablinger:2012qm,Ablinger:2015tua}, 
e.g. the Appell $F_1$ function. It has the integral representation
\begin{eqnarray}
F_1(a,b_1,b_2,c;x,y) &=& \frac{\Gamma(c)}{
\Gamma(a) \Gamma(c-a)} \int_0^1 dt~t^{a-1} (1-t)^{c-a-1} (1-xt)^{-b_1} (1-yt)^{-b_2}, 
\nonumber\\ &&
\hspace*{7.1cm} {\sf Re}(c) >
{\sf Re}(a) > 0.
\end{eqnarray}
All the  $_{p+1}F_p$ functions have a single infinite sum representation, while the Appell-functions
are represented by two infinite sums. Meeting these structures a larger number of integrals is mapped to
these remaining sums only.
In the physical applications the parameters $a_i, b_i,c_i, ...$
are functions of the dimensional parameter $\varepsilon$, in which these quantities have to be expanded.
One finally obtains infinite sum representations, which have to be calculated using general summation 
methods, cf.~Section~\ref{BS:sec:37}. There are also some other classes of named higher transcendental 
functions, which obey multi-sum representations \cite{EXTON1,EXTON2,SRIKARL}. For more complicated 
integrand-structures, however, one has to apply other techniques of integration, to which we turn now.
\subsection{Mellin-Barnes integrals}
\label{BS:sec:34}

\vspace*{1mm}
\noindent
All the higher transcendental functions being discussed in Section~\ref{BS:sec:33} have representations
in terms of Pochhammer-Umlauf integrals, see \cite{HYPKLEIN}, and also by Mellin-Barnes integrals.
It provides also the natural generalization for cases in which the representation in terms of special
functions is not known. Feynman parameter integrals usually exhibit denominators which are 
hyper-exponential, i.e. are products of factors $(P_j(x_i))^{\lambda_j},~\lambda_j \in \mathbb{R}, \lambda_j 
> 0$. One may apply the Mellin-Barnes decomposition (\ref{eq:MB}) to them. Here one may use the 
packages and methods 
described in~\cite{Czakon:2005rk,Smirnov:2009up,Gluza:2007rt,Gluza:2010rn,Dubovyk:2016ocz,Dubovyk:2016aqv}. 
As outlined above, one obtains 
nested infinite sum representations after applying the residue theorem. These sums have to be solved
by applying summation techniques, see Section~\ref{BS:sec:37}. On the other hand, one may obtain 
numerical results also in performing the  Mellin-Barnes integrals numerically, cf. e.g. 
\cite{Gluza:2007rt,Gluza:2010rn,Dubovyk:2016ocz,Dubovyk:2016aqv}, which provides useful 
checks. In aiming at an analytic result one should introduce Mellin-Barnes representations as little 
and as careful as possible, to avoid to shift the solution of the problem widely to the summation 
techniques.
\subsection{The method of hyperlogarithms}
\label{BS:sec:35}

\vspace*{1mm}
\noindent
If a Feynman diagram has no pole terms in the dimensional parameter $\varepsilon$ or can be made 
finite by certain transformations splitting off its pole terms \cite{vonManteuffel:2014qoa}, it can be 
calculated under certain conditions using the method of hyperlogarithms \cite{Brown:2008um}.
Since here the denominator of the Feynman integral is a multinomial in the Feynman parameters 
$x_i \in [0,1]$ one may seek a sequence of integrations, such that the denominator always is
a linear function (Fubini sequence) in the integration variable under consideration. In this case the 
Feynman integral can be found as a linear combination of Kummer-Poincar\'{e} iterated integrals, see 
Section~\ref{BS:sec:43}. The method has been first devised for massless scalar integrals in 
\cite{Brown:2008um}, for a corresponding code see \cite{Panzer:2014caa}. The method has been generalized
to massive Feynman diagrams, also containing local operator insertions in Ref.~\cite{Ablinger:2014yaa},
extending its validity even to cases with no thorough multi-linearity. An implementation can be found in
\cite{Wissbrock:2015faa}.
\subsection{The Almkvist-Zeilberger Algorithm}
\label{BS:sec:36}

\vspace*{1mm}
\noindent
Feynman parameter integrals usually appear as multi-integrals over $\{x_i|_{1=1}^n\} \in [0,1]^n$.
Furthermore, in the single-variate case, they depend on a real variable $x$ as $I(x)$ or an integer 
variable
$N$ as $I(N)$, and usually will depend in addition on the dimensional parameter $\varepsilon$.
One is now interested in finding either an associated differential equation for $I(x)$ or a difference
equation for $I(N)$. The Almkvist-Zeilberger algorithm \cite{AZ1,AZ2} allows to find these equations
which are of the form
\begin{eqnarray}
\label{eq:AZ1}
\sum_{l=0}^m P_l(x,\varepsilon) \frac{d^l}{dx^l} I(x,\varepsilon) &=& N(x,\varepsilon)
\\
\label{eq:AZ2}
\sum_{l=0}^m R_l(N,\varepsilon) I(N+l,\varepsilon) &=& M(N,\varepsilon),
\end{eqnarray}
where $P_l, R_l$ are polynomials and $N(x,\varepsilon), M(N,\varepsilon)$ denote potential
inhomogeneities. Internally, the equations (\ref{eq:AZ1}, \ref{eq:AZ2}) are found by solving an underlying linear system of equations.
The algorithm transforms an $n$-fold integration problem into a single equation, which can then be 
solved, by applying the techniques described in the subsequent sections. The bounds on $m$, given in the 
literature, are usually much higher, than what is found in praxis.
An optimized and improved algorithm for the input class of Feynman integrals has 
been implemented in the \texttt{MultiIntegrate} package~\cite{Ablinger:2013hcp,Ablinger:2015tua}. It can either produce equations of the form~\eqref{eq:AZ1},\eqref{eq:AZ2} where the right hand sides are zero, or it can provide inhomogeneous equations with further tactics to simplify the inhomogeneous sides using tools introduced in the next subsections.
\subsection{Differential equations}
\label{BS:sec:38}

\vspace*{1mm}
\noindent
The reduction of Feynman integrals to master integrals by the IBP-relations may also be used to
obtain a system of ordinary differential equations in one of the parameters of the master integrals,
or a system of partial differential equations in the case of more parameters. The master integrals 
may then be obtained as the solution of these systems under appropriate boundary conditions,  
\cite{Kotikov:1990kg,Bern:1992em,Remiddi:1997ny,Gehrmann:1999as}. For single-variate systems one considers
\begin{eqnarray}
\label{eq:DEQ1}
\frac{d}{dx} 
\left(
\begin{array}{c}
f_1\\ \vdots \\ f_n\end{array}\right) 
= \left(\begin{array}{ccc} 
A_{11} & \hdots & A_{1,n}
\\ 
\vdots &  & \vdots 
\\
A_{n1} & \hdots & A_{n,n}
\end{array} \right)
\left(\begin{array}{c}f_1\\ \vdots \\ f_n\end{array}\right) 
+ \left(\begin{array}{c}g_1\\ \vdots \\ g_n\end{array}\right),
\end{eqnarray}
which may be transformed into the scalar differential equation
\begin{eqnarray}
\label{eq:DEQ2}
\sum_{k=0}^n p_{n-k}(x)\frac{d^{n-k}}{dx^{n-k}} f_1(x) = \overline{g}(x),
\end{eqnarray}
with $p_n\neq0$, and $(n-1)$ equations for the remaining solutions, which are fully determined by the solution $f_1(x)$.

An important class is formed by the 1st order factorizing systems, cf.~Section~\ref{BS:sec:2}, after applying 
the decoupling methods \cite{Zuercher:94,Ablinger:2013jta} encoded in {\tt Oresys} \cite{ORESYS}. They appear
as the most simple case. One may transform the decoupled system into Mellin space and use the efficient solvers
of the package {\tt Sigma} there, cf.~Ref.~\cite{Ablinger:2015tua} and Section~\ref{BS:sec:37}.

The decoupled differential operator of (\ref{eq:DEQ2}) can be written in form of a 
non-commutative product of first order differential operators and rational terms, which
can now be integrated directly. The result is given by iterative integrals, also using partial 
fractioning and integration by parts to obtain representations in terms of proper letters,
\begin{eqnarray}
\label{eq:DEQ4}
f_1(x) &=& \sum_{k = 1}^{n+1} \gamma_k g_{k}(x),~\gamma_k \in {\mathbb C},\\
g_{k}(x) &=& h_0(x)\int_0^x dy_1 h_{1}(y_1) \int_0^{y_1} dy_2 h_{2}(y_2) ... 
\int_0^{y_{k-2}} dy_{k-1} h_{k-1}(y_{k-1})\int_0^{y_{k-1}} dy_{k} q_k(y_{k})
\end{eqnarray}
with $q_k(x)=0$ for $1\leq k\leq m$. Further, $\gamma_{m+1}=0$ if $\bar{g}(x)=0$ 
in~\eqref{eq:DEQ2}, and $\gamma_{m+1}=1$ and $q_{m+1}(x)$ being a mild variation of 
$\bar{g}(x)$ if $\bar{g}(x)\neq0$.
Note that these solutions are also called d'Alembertian solutions~\cite{Abramov:94}.
As the master integrals appearing in quantum field theories obey differential equations with rational 
coefficients, the letters $h_i$, which constitute the iterative integrals, have to be algebraic, see 
Section~\ref{BS:sec:4}. Given a differential equation~\eqref{eq:DEQ2}, the d'Alembertian solutions~\eqref{eq:DEQ4} can be computed with the package~\texttt{HarmonicSums}~\cite{Ablinger:2017Mellin}. More generally, also Liouvillian solutions~\cite{Singer:81} can be calculated with~\texttt{HarmonicSums} utilizing Kovacic's algorithm~\cite{Kovacic:86}; this algorithms finds also Liouvillian solutions of second order differential equations.
Summarizing, representation (\ref{eq:DEQ4}) can be found in the 1st order decoupling case of a coupled system of the form~\eqref{eq:DEQ1} for 
any 
basis 
of master integrals, cf. \cite{Ablinger:2015tua}\footnote{This algorithm has been applied in many massive 
3-loop calculations so far, cf. \cite{Ablinger:2017ptf} for a survey, and also in \cite{Ablinger:2017hst}.}. 
If being transformed to the associated system of difference equations, the same 
holds, if this system is also first order factorizing. The solution of the remaining equations are directly obtained
by the first solution.

The corresponding algorithms hold to 
all orders in $\varepsilon$, on the expense that it is necessary to obtain higher order expansions to cancel the 
respective poles in the expansion of the coefficients. This algorithm has some relation to the method of 
hyperlogarithms, cf.~Section~\ref{BS:sec:35}. For an implementation see Ref.~\cite{FORMF3}. For the solution
of 1st order decoupling systems also algorithms given in \cite{Raab:2012} are useful.

In the multi-variate case, the $\varepsilon$-representation of a linear system of partial differential equations 
\begin{eqnarray}
\label{eq:DEQ5}
\partial_m f(\varepsilon, x_n) = A_m(\varepsilon, x_n) f(\varepsilon, x_n)
\end{eqnarray}
is important, as has been recognized in  Refs.~\cite{Kotikov:2010gf,Henn:2013pwa}, see also \cite{Henn:2014qga}.
One considers the well-known zero-curvature representations \cite{NOVIKOV:80,Sakovich:1995}
\begin{eqnarray}
\partial_n A_m - \partial_m A_n + [A_n,A_m] = 0,
\end{eqnarray}
with $[A,B]$ a Lie bracket \cite{SLIE}. The matrices $A_n$ can now be transformed (in the general non-Abelian case) 
by
\begin{eqnarray}
A_m' = B^{-1} A_m B - B^{-1}(\partial_m B),
\end{eqnarray}
see also \cite{NOVIKOV:80,Sakovich:1995}. One now intends to find a matrix $B$ to transform (\ref{eq:DEQ5}) into the form
\begin{eqnarray}
\label{eq:DEQ6}
\partial_m f(\varepsilon, x_n) = \varepsilon A_m(x_n) f(\varepsilon, x_n),
\end{eqnarray}
if possible. This then allows solutions in terms of iterative integrals. The corresponding system 
(\ref{eq:DEQ5}),
therefore, has also to belong to the first order factorizing case. A formalism for the basis change to the 
$\varepsilon$-basis has 
been proposed in \cite{Lee:2014ioa} and implemented in the single-variate case in 
\cite{Prausa:2017ltv,Gituliar:2017vzm} and in 
the multi-variate case in \cite{Meyer:2017joq}. In this description one has the advantage to work at equal weight, 
which avoids necessary deeper expansions in $\varepsilon$ needed otherwise.
\subsection{The method of arbitrarily high moments}
\label{BS:sec:39}

\vspace*{1mm}
\noindent
Standard procedures like {\tt Mincer} \cite{Gorishnii:1989gt}\footnote{
See also \cite{Baikov:1996rk,Smirnov:2003kc} for Baikov's method, which allowed a large number
of calculations up to the 5-loop level as e.g. \cite{Baikov:2016tgj}.
For the solution of important 4-loop 
problems {\tt Forcer} \cite{Ruijl:2017cxj} and the $R^*$ \cite{Herzog:2017bjx}
method have been developed.}, {\tt MATAD} \cite{Harlander:1997zb} or {\tt Q2E} 
\cite{Harlander:1997zb,Seidensticker:1999bb} allow the calculation of a comparable small number of Mellin 
moments of a given quantity only, because the resources needed by these algorithms grow exponentially.
Alternatively, one may think of using difference equations for the master integrals of a physical problem in 
Mellin space to calculate the integer moments of the physical quantity studied to obtain higher and higher 
moments. This has been pursued in Ref.~\cite{Blumlein:2017dxp} and the formalism also applies when only ordinary
differential equations are available for the master integrals. Furthermore, here we need not to know the degree
of decoupling of the corresponding system, which can be rather high.

The most extensive system having been studied using this method so far led to 8000 moments \cite{Ablinger:2017ptf} 
after several weeks of computation time. In applications the structure of the Mellin moments is usually 
mathematically much simpler than the corresponding general expressions in $x$ and $N$ space. They are 
given by rational numbers and are labeled by special constants, like MZVs. Furthermore, the physical quantity to 
be calculated can turn out to be structurally less complex than the master integrals. Moreover, unlike the latter,
one has to calculate it only to $O(\varepsilon^0)$. 

Finally, one projects onto color factors and special constants to obtain rational sequences. They are used as 
input to guessing to seek for a closed form difference equation in all cases. We could achieve this 
also in all the complex cases mentioned. The final 
step consists now in solving these difference equations using the methods of the package {\tt Sigma}. For a 
series of difference equations a solution has been found. In others only the 1st order factorizing factors were 
obtained, leaving remainder 4th order systems, which are further investigated using the methods described in 
Section~\ref{BS:sec:5}. Here the last system decoupled into two $2 \times 2$ systems.

The method of arbitrarily high moments is a rather global one. E.g. in 1st order factorizable systems like they 
appear in massless calculations, e.g. for the 3 loop anomalous dimensions \cite{Ablinger:2017tan} and related 
quantities, the method is fully automated.
\subsection{Summation Techniques in \boldmath $\Pi\Sigma$-Fields and Rings}
\label{BS:sec:37}

\vspace*{1mm}
\noindent  
As described in Sections~\ref{BS:sec:33} and~\ref{BS:sec:36}, the evaluation of Feynman integrals can be often 
reduced to the problem to simplify multiple sums defined over hypergeometric products, harmonic numbers and 
generalizations 
thereof~\cite{Ablinger:2013cf,Ablinger:2013jta,Ablinger:2013eba,Blumlein:1998if,Ablinger:2011te,Ablinger:2014bra}; 
see also Section~\ref{BS:sec:4}. Given such a sum representation, symbolic summation algorithms in the setting 
of difference rings and fields~\cite{Karr:1981,Bron:00,Schneider:01,Schneider:04a,Schneider:05a,Schneider:05b,
Schneider:07d,Schneider:10b,Schneider:10c,Schneider:15a,Schneider:08c,Schneider:08d,Schneider:08e} can be 
utilized to derive an alternative representation in terms of indefinite nested sums defined over hypergeometric 
products. Special cases of these sums are enumerated in Section~\ref{BS:sec:4}. More precisely, using the 
Mathematica package \SigmaP~\cite{Schneider:2007a,Schneider:2013a} the following three steps lead often to the 
desired simplification: (1) compute a linear recurrence relation for the sum using the creative telescoping 
paradigm~\cite{AequalB} in the setting of difference rings, and (2) solve afterwards the recurrence in terms of 
d'Alembertian solutions~\cite{Abramov:94,Petkov:2013}. This means that one finds all solutions that can be given in terms indefinite nested sums defined over hypergeometric products. Note that all the sums from Section~\ref{BS:sec:4} are covered.
Finally, in step (3) one tries to combine the solutions to derive a simpler representation of the original input sum.
The package \texttt{EvaluateMultiSums}~\cite{Ablinger:2010pb,Schneider:2013zna} based on \SigmaP\ combines all these steps and variants thereof in a very efficient way to carry out such simplifications automatically. Here one works from inside to outside: one considers first the innermost definite sum and transforms it with the above method to an expression in terms of indefinite nested sums. Afterwards, one treats the next definite summation quantifier and transforms this sum to an expression in terms of indefinite nested sums, etc.

A crucial step for the above approach is that in a preprocessing step one first gets rid of the dimensional parameter $\ep$ by 
expanding first the summand of the multiple sum w.r.t.\ $\ep$, i.e., computes the first coefficients of its Laurent series expansion and by applying afterwards the summation quantifiers to each of the coefficients. Only then the summation technologies described above are applied to the summations of the coefficients up to the desired order. Alternatively, there are algorithms available that can compute directly for a multiple sum a recurrence of the form~\eqref{eq:AZ2} depending also on $\ep$. In particular, if the summand consists of a product of hypergeometric products, one can utilize WZ-summation theory~\cite{Wilf:92}; for refined methods and an efficient implementation we refer to the \texttt{MultiSum} package~\cite{Wegschaider}. This approach can be considered as the discrete version of the Almkvist-Zeilberger algorithm introduced in Section~\ref{BS:sec:36}.  Note that one can use also holonomic summation algorithms~\cite{Zeilberger:90a,Chyzak:00} to derive recurrences of the form~\eqref{eq:AZ2}; for a generalized version combining the holonomic and difference ring approach we refer to~\cite{Blumlein:2017ztz} and references therein. 
Finally, one can use \texttt{Sigma}'s recurrence solver described in~\cite{Blumlein:2010zv}.
Suppose that one is given a recurrence of the form~\eqref{eq:AZ2} where the inhomogeneous part is given as an $\ep$-expansion; more precisely, the first coefficients are given in terms of indefinite nested sums defined over hypergeometric products. Then together with the first initial values of the Feynman integral $I(N)$, which is a solution of~\eqref{eq:AZ2}, one can decide constructively if the coefficients of the $\ep$-expansion of $I(N)$ can be again given in terms of indefinite nested sums defined over hypergeometric products. Note that \texttt{Sigma}'s recurrence solver can be also utilized to solve recurrences coming, e.g., from the Almkvist-Zeilberger algorithm; see Section~\ref{BS:sec:36}.

In many applications, the transformation of Feynman integrals yields an enormous expression consisting of 
up to thousands of multiple sums. To simplify these sums successively using these summation tools is not 
only problematic because of time restrictions, but also because of the following intrinsic problem: 
The 
scattered sums themselves cannot be simplified in terms of indefinite nested sums, but only a suitable combination of them can be simplified in this way. In order to bypass these problems, the package \texttt{SumProduction}~\cite{Blumlein:2012hg,Schneider:2013zna} built on \texttt{Sigma} can be utilized: it reduces the sum expressions to compact forms where the arising sums are merged appropriately. Afterwards \texttt{EvaluateMultiSum} can be applied to this optimized expression within reasonable time and without dealing with sums that cannot be handled within the difference 
ring setting.

Besides the simplification of multiple sums, the difference ring approach can be also used to solve coupled 
systems of differential and difference equations~\cite{Bluemlein:2014qka,Ablinger:2015tua,CoupledSys:15}. 
More precisely, given a finite number of Feynman integrals $F_1(N),\dots,F_n(N)$ depending on a Mellin 
variable $N$ which are determined by a system of linear difference equations equipped with sufficiently 
many initial conditions, the package \texttt{SolveCoupledSystem} decides algorithmically if the Feynman 
integrals can be expressed in terms of indefinite nested sums defined over hypergeometric products. Internally, 
the following machinery is applied. (1) Z\"urcher's algorithm~\cite{Zuercher:94} implemented in the package 
\texttt{OreSys}~\cite{ORESYS} is used to decouple the system to one scalar recurrence that characterizes one 
of the given Feynman integrals, say $F_1(N)$, together with explicitly given linear combinations of $F_1(N)$ 
(and possible shifts) describing the remaining integrals $F_2(N),\dots,F_n(N)$ (sometimes one obtains several 
scalar recurrences for a subset of the given integrals). (2) Then using \texttt{Sigma}'s recurrence solver one 
can decide constructively if the Feynman integral $F_1(N)$ (or the coefficients of the $\ep$-expansion of $F_1(N)$) 
can be written in terms of indefinite nested sums. If this is not possible, it follows that also $F_2(N),\dots,F_n(N)$ 
cannot be expressed in terms of indefinite nested sums. However, if it is possible, also $F_2(N),\dots,F_n(N)$ can be 
assembled to expressions in terms 
of indefinite nested sums using the linear combination of $F_1(N)$ (and its shifted versions).\\
In addition, this method can be generalized to solve coupled systems of differential equations coming from IBP 
methods~\cite{Chetyrkin:1981qh,Laporta:2001dd,Studerus:2009ye,vonManteuffel:2012np,MARSEID}; see also 
Sections~\ref{BS:sec:2}, \ref{BS:sec:38}. Here one assumes that the Feynman integrals $f_1(x),\dots,f_n(x)$ can be 
given in the power series representations
$$f_i(x)=\sum_{N=0}^{\infty}F_i(N) x^N$$
for $i=1,\dots,n$. Then given a coupled system of linear differential equations of the form\footnote{In addition, we assume that the the inhomogeneous parts $g_i(x)$ in~\eqref{eq:DEQ1} are given in power series representations were the coefficients can be written in terms of indefinite nested sums defined over hypergeometric products.}~\eqref{eq:DEQ1} in the $f_1(x),\dots,f_n(x)$, 
one can decide constructively, if the coefficients $F_1(N),\dots,F_n(N)$ (or the coefficients of 
their $\ep$-expansions) can be represented in terms of indefinite nested sums. Internally, this problem can be 
reduced to the problem to solve a coupled system of difference equations as follows. Comparing coefficients 
w.r.t.\ $x^N$, the coupled system of differential equations in the $f_1(x),\dots,f_n(x)$ yield a coupled system of linear difference equations in the $F_1(N),\dots,F_n(N)$, and afterwards the above algorithm is applied; 
compare~\cite{Bluemlein:2014qka,Ablinger:2015tua,CoupledSys:15}. Alternatively, one can first decouple 
the coupled system of linear differential equations, afterwards transforms the scalar differential 
equation to a scalar recurrence by comparing coefficients w.r.t.\ $x^N$, and finally one applies \SigmaP\ 
to solve the scalar recurrence. For further details we refer to~\cite{Schneider:2016szq}.
\section{Factorizing Corrections: The Structure Function Method}
\label{BS:sec:F}

\vspace*{1mm}
\noindent
In the case of radiation off (meta)stable massive fermions, like the charged leptons and heavy quarks
of mass $m$, the logarithmically enhanced contributions can be determined by the renormalization group 
equations
\cite{Symanzik:1970rt,Callan:1970yg} describing the factorization of the process of the corresponding 
scattering cross section 
\begin{eqnarray}
\label{eq:SF1}
\sigma(s) = C\left(\frac{s}{\mu^2}\right) \otimes A\left(\frac{\mu^2}{m^2}\right), 
\end{eqnarray}
neglecting power corrections. Here $s$ denotes the CMS energy squared and $\otimes$ the Mellin 
convolution
\begin{eqnarray}
A(x) \otimes B(x) = \int_0^1 dx_1 \int_0^1 dx_2 \delta(x - x_1 x_2) A(x_1) B(x_2).
\end{eqnarray}
One calls the corrections due to the function $A(x)$ factorizing, because of its connection to $C(x)$
by a Mellin convolution, which factorizes, if considered in Mellin space, via
\begin{eqnarray}
\Mvec[A(x) \otimes C(x)](N) = \Mvec[A(x)](N) \cdot \Mvec[C(x)](N).
\end{eqnarray}
Both the coefficient function $C(x)$ and the operator matrix elements $A(x)$
obey a series expansion in the coupling constant $a = \alpha/(4\pi)$,
\begin{eqnarray}
A(x)  &=& \sum_{k=0}^\infty a^k \sum_{l=0}^k A_{k,l}(x) \ln^l\left(\frac{\mu^2}{m^2}\right)
\\
C(x)  &=& \delta(1-x) + \sum_{k=1}^\infty a^k \sum_{l=0}^k C_{k,l}(x) \ln^l\left(\frac{s}{\mu^2}\right),
\end{eqnarray}
such that in the expansion of $\sigma(s)$ the dependence on the factorization scale $\mu$ vanishes and all
logarithmic terms depend on the ratio $s/m^2$ only, see e.g.~\cite{Behring:2014eya}. 
Here the expansion coefficients of the coefficient function $C_{k,l}$ are process dependent, while
the ones of the operator matrix elements $A_{k,l}$ are process independent. Since the decomposition
(\ref{eq:SF1}) has been studied for deep-inelastic structure functions first one calls this representation
also the structure function method.

Let us consider the QED corrections in the inclusive case as an example. The leading order corrections 
are of $O((a \ln(s/m^2))^k)$. These are universal. For the QED corrections, initial- and final state
corrections have been calculated to various processes, including light fermion emission, in 
Refs.~\cite{Jezabek:1991bx,Skrzypek:1992vk,Przybycien:1992qe,Blumlein:1996yz,
Arbuzov:1999cq,Blumlein:2004bs,Blumlein:2007kx} up to $O((a \ln(s/m^2))^5)$ in the unpolarized and 
polarized case. One may as well treat sub-leading logarithms, as lined out in 
Refs.~\cite{Berends:1987ab,Blumlein:2011mi,Blumlein:2002fy}, see also \cite{Blumlein:2012bf}. These corrections 
are no longer universal, 
but depend on the different contributing subprocesses of the observable considered. The structure function
method only applies to situations in which power corrections of $O(m^2/s)$ can be safely neglected. 
\section{Function Spaces for First Order Factorizing Systems}
\label{BS:sec:4}

\vspace*{1mm}
\noindent
When solving first order factorizing systems it is not known a priori, which particular 
alphabet will span the function (distribution) space of the solutions. All what is known
is the structure of the coefficients of the respective determining differential or difference 
equation and that the letters of the alphabet will result from them. In Section~\ref{BS:sec:3} 
we have outlined a series of solution algorithms which finally also determine the corresponding
alphabets. In the following we will concentrate on the cases which have been found in quantum field theory
so far. Usually the letters appearing in the different subsets may even appear combined in 
applications\footnote{For previous surveys see Refs.~\cite{Ablinger:2013jta,Ablinger:2013eba}.}. 
All the 
functions described in the following and their mutual relations are implemented in the package {\tt 
HarmonicSums} \cite{Vermaseren:1998uu,Blumlein:1998if,
Ablinger:2014rba, Ablinger:2010kw, Ablinger:2013hcp, Ablinger:2011te, Ablinger:2013cf, Ablinger:2014bra,Ablinger:2017Mellin}.
\subsection{Harmonic sums and harmonic polylogarithms}
\label{BS:sec:41}

\vspace*{1mm}
\noindent
Next to purely rational functions in either the Mellin variable $N$ or the momentum-fraction variable $x \in 
[0,1]$, one obtains the nested harmonic sums \cite{Vermaseren:1998uu,Blumlein:1998if}
\begin{eqnarray}
\label{eq:hsum}
S_{b,\vec{a}}(N) = \sum_{k=1}^N \frac{({\rm sign}(b))^k}{k^{|b|}} S_{\vec{a}}(k),~~~ a_i,b,N \in \mathbb{N} 
\backslash \{0\},~~ S_\emptyset = 1,
\end{eqnarray}
and the harmonic polylogarithms \cite{Remiddi:1999ew}
\begin{eqnarray}
\label{eq:hpol}
H_{b,\vec{a}}(x) = \int_0^x dy f_b(y) H_{\vec{a}}(y),~~~ b, a_i \in \{-1,0,1\},~~H_\emptyset(x) =1,
\end{eqnarray}
with the letters
\begin{eqnarray}
\label{eq:hpllett}
f_{-1}(x) = \frac{1}{1+x},~~~
f_{0}(x) = \frac{1}{x},~~~
f_{1}(x) = \frac{1}{1-x}.
\end{eqnarray}
There are subsets of (\ref{eq:hsum}), like the non-alternating harmonic sums with $a_i, b > 0$, and of 
(\ref{eq:hpol}) with the alphabets $\{0,1\}$ and $\{0,-1\}$, the Nielsen integrals 
\cite{NIELSEN,KMR70,Kolbig:1983qt,Devoto:1983tc}. A subset of the latter are the canonical polylogarithms 
\cite{LEWIN1,LEWIN2}.\footnote{Some integration methods may lead to very sophisticated argument 
structures of polylogarithms, harmonic polylogarithms and related functions. The method of the symbol 
\cite{Duhr:2011zq} can be used to simplify, and to essentially compactify the corresponding expressions.}

Both function spaces form shuffle algebras in the case of the functions, 
cf. \cite{REUTENAUER,Blumlein:2003gb}, and quasi-shuffle 
algebras in case of the sums, cf.~\cite{REUTENAUER,HOFFMAN,Blumlein:2003gb}\footnote{The property to form a 
shuffle algebra is quite general and it will apply also to the other function spaces discussed below.}, 
\begin{eqnarray}
\label{eq:SHUF1}
S_{a}(N) &\shuffle& S_{b_1,..., b_m}(N) = S_{a,b_1,..., b_m}(N) 
+ S_{b_1,a,..., b_m}(N) + ... + S_{b_1,...,b_m,a}(N) 
\\ 
\label{eq:SHUF2} H_a(x) \cdot H_{b_1, ..., b_m}(x) &=& 
H_a(x) \shuffle H_{b_1, ..., b_m}(x) = H_{a,b_1,...,b_m}(x) + H_{b_1,a,...,b_m}(x) 
+ ... + H_{b_1,..., b_m,a}(x) 
\nonumber\\ 
\end{eqnarray} 
and form Hopf-algebras \cite{HOPF1,HOPF2,HOPF3},
which also form the basis of renormalizable quantum field theories 
\cite{Kreimer:1997dp,Connes:1998qv,Connes:1999yr,Connes:2000fe}. The bases of these algebras can be 
counted using Lyndon words 
\cite{LYNDON,RADFORD} and by using Witt formulae \cite{WITT1,WITT2}. Sum products obey quasi-shuffle relations 
as
\begin{eqnarray}
\label{eq:SHUF3}
S_{a}(N) \cdot S_{b_1,..., b_m}(N) &=& 
S_{a}(N) \shuffle S_{b_1,..., b_m}(N) 
+ S_{a_1 \wedge b_1,b_2,...,b_m}(N) 
+ S_{b_1,a_1 \wedge b_2,b_3,...,b_m}(N) 
\nonumber\\ &&
+ ...
+ S_{b_1,...,b_{m-1},a_1 \wedge b_m}(N), 
\end{eqnarray} 
with $a \wedge b = {\rm sign}(a) {\rm sign}(b)(|a|+|b|)$. It has been shown in~\cite{AS:18} that the existing relations given in the quasi-shuffle algebra coincide with the relations in the ring of sequences when one evaluates the elements to sequences.
To both spaces there is a single space of special constants associated, 
which is obtained in the limit $N 
\rightarrow \infty$ in case of the sums and $x \rightarrow 1$, provided the limits exist. These are the multiple 
zeta values (MZVs) \cite{Blumlein:2009cf}. There are more relations between these constants, 
than between the nested sums
and the iterated integrals. For the alphabets $\{0,1\}$ the relations were given up to {\sf w = 22} and
for $\{-1,0,1\}$ to {\sf w = 12}, respectively, in Ref.~\cite{Blumlein:2009cf}. One may add the symbol
$\sigma_0 := \sum_{k=1}^\infty (1/k)$ to the set formally, expressing the respective degree of divergence. For 
the set $\{-1,0,1\}$ the basic constants through polynomials of which all other constants can be obtained are
\begin{eqnarray}
\label{eq:MZVbas}
\left\{\sigma_0, \ln(2);
~\zeta_2;
~\zeta_3;
~\Li_4\left(\frac{1}{2}\right);
~\zeta_5,\Li_5\left(\frac{1}{2}\right);
~\Li_6\left(\frac{1}{2}\right);
~\zeta_7,\Li_7\left(\frac{1}{2}\right),\sigma_{-5,1,1},
\sigma_{5,-1,-1}; ...\right\}
\end{eqnarray}
from weight {\sf w = 1} to {\sf 7}. Here 
\begin{eqnarray}
\label{eq:MZVbas1}
\zeta_k = \sum_{l=1}^\infty \frac{1}{l^k},~~k \in \mathbb{N}, k \geq 2;~~~~
\Li_k(x) = \sum_{l=1}^\infty \frac{x^l}{l^k},~|x| \leq 1;~~~~
\sigma_{a,b,c} = \lim_{N \rightarrow \infty} S_{a,b,c}(N).
\end{eqnarray}

The Mellin transform maps the elements of both spaces, 
i.e. the Mellin transform of a harmonic polylogarithm can be represented by a linear combination of nested 
harmonic sums and their values in the limit $N \rightarrow \infty$, e.g.
\begin{eqnarray}
\label{eq:ex1}
\Mvec[H_{0,1,1}(x)](N) = - \frac{S_{1,1}(N)}{N^2} + \frac{\zeta_3}{N}~.
\end{eqnarray}
Applying the quasi-shuffle relations one may further reduce
\begin{eqnarray}
\label{eq:ex2}
S_{1,1}(N) = \frac{1}{2} \left[S_1^2(N) + S_2(N) \right]~.
\end{eqnarray}
The harmonic sums can be analytically continued from $N \in \mathbb{N}$ to $N \in \mathbb{C}$ for even resp. odd 
moments using their representation in terms of Mellin transforms. The latter are expanded into factorial series
\cite{FACT1,FACT2} and factors of $S_{1, ..., 1}(N)$, given by polynomials of poly-gamma functions and 
MZVs. Due to this 
one may differentiate harmonic sums for $N$ and derive besides the quasi-shuffle relations structural relations
\cite{Blumlein:2009fz,Blumlein:2009ta}. Approximate numerical representations for the analytic continuation of 
harmonic sums have also been obtained \cite{Blumlein:2000hw,Blumlein:2005jg,Kotikov:2005gr}. 
It finally can be shown that the inclusive 2-loop results in the massless and single-mass case, 
neglecting power 
corrections, can be expressed by only {\it six} harmonic sums \cite{Blumlein:2005im,Blumlein:2006rr}
\begin{eqnarray}
\label{eq:SIX}
S_1(N),~ 
S_{2,1}(N),~ 
S_{-2,1}(N),~ 
S_{-3,1}(N),~ 
S_{2,1,1}(N),~ 
S_{-2,1,1}(N),
\end{eqnarray}
see also \cite{Blumlein:2009fz,Blumlein:2009ta}. Harmonic sums and polylogarithms are also implemented in the 
packages {\tt summer} \cite{Vermaseren:1998uu}, {\tt harmpol} \cite{Remiddi:1999ew} 
and {\tt HPL} \cite{Maitre:2005uu}.
Precise numerical representations of harmonic sums were given in 
\cite{Gehrmann:2001pz,Ablinger:2017tqs}.
\subsection{Cyclotomic harmonic sums and harmonic polylogarithms}
\label{BS:sec:42}

\vspace*{1mm}
\noindent
An extension of the harmonic sums and polylogarithms are the cyclotomic harmonic sums and harmonic 
polylogarithms, the properties of which have been studied in Ref.~\cite{Ablinger:2011te}. The cyclotomic
harmonic polylogarithms are the iterated integrals over the alphabet
\begin{eqnarray}
\label{eq:cycl1}
\left\{\frac{1}{x}, \left. \frac{x^l}{\Phi_k(x)}\right|_{0 \leq l \leq \varphi(k)}\right\},
\end{eqnarray}
with $\Phi_k(x)$ the $k$th cyclotomic polynomial consecutively found by factorizing $(x^N -1)$ and 
$\varphi(k)$ the totient function. Iterated integrals of these letters are called 
cyclotomic harmonic polylogarithms, 
where $\Phi_1(x) = 1/(1-x), \Phi_2(x) = 1/(1+x), \Phi_3(x)=1/(1+x+x^2), 
\Phi_4(x)=1/(1+x^2)$ and $\Phi_6(x) = 1/(1-x+x^2)$. 
The Mellin transform of $x/(1-x +x^2)$ for $n = 6N$ is given by
\begin{eqnarray}
\label{eq:cycl2}
\Mvec\left[\frac{x}{1-x+x^2}\right](6N) &=& 
\frac{\big(4+38 N-47 N^2+127 N^3+90 N^4-1404 N^5-1080 N^6\big)}
{12 N (2N-1) (2N+1) (3N-1) (3N+2) (6N-5) (6N+1)} 
\nonumber\\ &&
+\frac{2 \pi }{3\sqrt{3}(6N-5)}
+ \frac{1}{6N-5} \Biggl[
  \frac{1}{3} \sum_{k=1}^N \frac{1}{1+2 k}
+ \frac{1}{2} \sum_{k=1}^N \frac{1}{2+3 k}
\nonumber\\ &&
+ \sum_{k=1}^N \frac{1}{1+6 k}
- \frac{1}{6} S_1(N)
\Biggr],
\end{eqnarray}
as an example. 
Here cyclotomic harmonic sums such as $\sum_{k=1}^N \frac{1}{2+3 k}$ emerge, i.e. harmonic sums
exhibiting a cyclic pattern of missing terms in their summands. 
New special constants are obtained from the cyclotomic harmonic sums in the limit $N 
\rightarrow \infty$ and the cyclotomic harmonic polylogarithms for $x \rightarrow 1$. One of these is 
Catalan's constant
\begin{eqnarray}
{\sf C} = \sum_{l=0}^\infty \frac{(-1)^l}{(2l+1)^2}.
\end{eqnarray}
Cyclotomic harmonic polylogarithms contribute to the massive form factors at 3-loop order
\cite{FORMF3,Henn:2016tyf}. Real-valued cyclotomic harmonic polylogarithms can be decomposed into 
complex-valued generalized harmonic polylogarithms by partial fractioning. Expressions based on the latter 
representations are often more complicated to deal with. One example is the significantly extended ground 
field in case of sum-representations of their Mellin transforms, which gives preference to the real 
representations.
\subsection{Generalized harmonic sums and harmonic polylogarithms}
\label{BS:sec:43}

\vspace*{1mm}
\noindent
Generalized harmonic polylogarithms were introduced as iterated integrals by Kummer \cite{KUMMER} and
also studied by Poincar\'{e} \cite{POINCARE}, see also \cite{LADAN,CHEN,GONCHAROV}. These functions 
emerged in higher loop calculations in \cite{Borwein:1999js,Moch:2001zr} and the properties of them have been
detailed in Ref.~\cite{Ablinger:2013cf}. One considers
\begin{eqnarray}
\label{eq:GHPL1}
G_{b,a_1,...,a_m}(x) = \int_0^x dy~g_b(y) G_{a_1,...,a_m}(y),~~~b, a_i \in \mathbb{C},~~~G_\emptyset = 1, 
\end{eqnarray}
with
\begin{eqnarray}
\label{eq:GHPL1a}
g_c(x) = \frac{1}{x-c}.
\end{eqnarray}
Any function $1/P(x)$, with $P(x)$ a polynomial with coefficients in $\mathbb{C}$ can be decomposed into letters
$g_c(x)$. Therefore, the framework of generalized harmonic polylogarithms is very general.
The Mellin transform of (\ref{eq:GHPL1}) leads to linear combinations of the associated nested sums,
\begin{eqnarray}
\label{eq:GHPL2}
S_{c,d_1,...,d_m}(b,a_1,...,a_m;N) = \sum_{k = 1}^N \frac{{b}^k}{k^c} S_{d_1,...,d_m}(a_1,...,a_m;k),
~~c, d_k \in \mathbb{N} \backslash \{0\}, b, a_i \in \mathbb{C},~~S_\emptyset = 1. 
\end{eqnarray}
Likewise, special associated numbers are obtained in the limits $N \rightarrow \infty, x \rightarrow 1$, if the 
limits exist. In single scale applications in quantum field theory the letters in (\ref{eq:GHPL1}) are 
often rational numbers, respectively roots of unity, in the cyclotomic case, however, due to general polynomial 
structures also other algebraic numbers may occur. In the presence of more than one scale they are general 
complex numbers. Calculating the massless 3-loop Wilson coefficients for deep-inelastic scattering, generalized 
harmonic sums appeared in intermediary steps but canceled in the finite results \cite{Vermaseren:2005qc}. This 
applies also to the case of the $N_F$-terms for the 3-loop massive OMEs \cite{Ablinger:2010ty,Blumlein:2012vq}. 
However, in the 3-loop pure-singlet massive OME \cite{Ablinger:2014nga} generalized harmonic sums contribute.
Relations for generalized harmonic polylogarithms are also implemented in the packages {\tt 
nestedsums}~\cite{Weinzierl:2002hv} and {\tt Xsummer} \cite{Moch:2005uc}. A numerical implementation
for generalized harmonic polylogarithms has been given in \cite{Vollinga:2004sn}.
\subsection{Binomial and inverse binomial sums and square-root valued iterated integrals}
\label{BS:sec:44}

\vspace*{1mm}
\noindent
At the next level of complexity, square-root valued letters of a quadratic form with integer coefficients for 
iterated integrals emerge. Examples are
\begin{eqnarray}
\label{eq:sq1}
\frac{1}{(2+x) \sqrt{x-\tfrac{1}{4}}},~~~~~~\frac{1}{(1+x) \sqrt{x}\sqrt{8-x}},
\end{eqnarray}
cf.~\cite{Ablinger:2014bra}. They are related to binomial and inverse binomial structures of the 
following kind 
\begin{eqnarray}
\label{eq:sq2}
\binom{2N}{N} = \frac{4^N}{\pi} \Mvec\left[\frac{1}{\sqrt{x(1-x)}}\right](N),~~~
\frac{1}{N \displaystyle \binom{2N}{N}} = \frac{1}{4^N} \Mvec\left[\frac{1}{x\sqrt{1-x}}\right](N)
\end{eqnarray}
in Mellin space. In Ref.~\cite{Ablinger:2014bra} a large number of these letters, characterized by different 
classes, have been studied. 

In general, the Mellin transform of the iterated integrals of these square-root valued letters are related 
to nested sums containing (inverse) binomial weights in front of generalized harmonic sums, like
\begin{eqnarray}
&& \hspace*{-4mm}
\sum_{i=1}^N \binom{2i}{i} (-2)^i \sum_{j=1}^i \frac{1}{\displaystyle j \binom{2j}{j}}
S_{1,2}\left(\frac{1}{2},1\right)(j)
=
\int_0^1 dx \frac{(-x)^N-1}{x+1}\sqrt{\frac{x}{8-x}}\Bigl[\HA_{\sf w_{12},1,0}(x)-2\HA_{\sf w_{13},1,0}(x)
\nonumber\\
&&
\hspace*{-4mm}
-\zeta_2\left(\HA_{\sf w_{12}}(x)-2\HA_{\sf w_{13}}(x)\right)\Bigr]
-\frac{5\zeta_3}{8\sqrt{3}}\int_0^1 dx \frac{(-2x)^N-1}{x+\frac{1}{2}}\sqrt{\frac{x}{4-x}}
+c_1\int_0^1 dx \frac{(-8x)^N-1}{x+\frac{1}{8}}\sqrt{\frac{x}{1-x}},
\nonumber\\
\\
&&
\hspace*{-4mm}
c_1=\frac{1}{\pi}\sum_{j=1}^\infty \frac{1}{\displaystyle j \binom{2j}{j}}
S_{1,2}\left(\frac{1}{2},1\right)(j) \approx 0.10184720\dots,
\end{eqnarray}
with $\HA_{b,\vec{a}}(x) = \int_x^1 dy f_b(y) \HA_{\vec{a}}(y)$. Algorithms to compute the (inverse) Mellin transform in terms of indefinite nested sums/integrals are available in \texttt{HarmonicSums}~\cite{Ablinger:2017Mellin}; note that these algorithms are based on linear recurrence and differential equation solving as described in Subsections~\ref{BS:sec:38} and~\ref{BS:sec:37}.
Infinite (inverse) binomial sums depending on one parameter $\xi \in \mathbb{C}$ have been also studied in
\cite{Davydychev:2003mv,Weinzierl:2004bn}. Finite binomial sums appear in various massive calculations, see e.g.
Refs.~\cite{Ablinger:2014uka,Ablinger:2015a,Ablinger:2015tua}. 

In this context also new special constants are appearing, as e.g.
\begin{eqnarray}
\label{eq:sq3}
{\rm arccot}(\sqrt{7}),~~~~~\Li_n\left(-\frac{1}{4}\right),
\end{eqnarray}
and many others.

One may think of to rationalize the root-valued letters in trying to map the problem to Kummer-Poincar{\'e} 
iterated integrals. This is indeed possible for a series of cases. Concerning all the cases discussed in 
Ref.~\cite{Ablinger:2014bra}, a successful transformation could, however, not be found, cf.~\cite{BR}.

Finally, we would like to mention, that some elliptic solutions in $x$ space are hypergeometric
in $N$-space. The Mellin transform
of the following elliptic integrals of the first and the second kind obey 
\cite{Ablinger:2013eba,Ablinger:2017bjx}
\begin{eqnarray}
\label{eq:Ksimp}
\Mvec[{\bf K}(1-z)](N)  = \frac{2^{4N+1}}{\displaystyle (1+2N)^2 \binom{2N}{N}^2},~~~
\Mvec[{\bf E}(1-z)](N)  = \frac{2^{4N+2}}{\displaystyle (1+2N)^2 (3+2N) \binom{2N}{N}^2}.
\end{eqnarray}
Eqs.~(\ref{eq:Ksimp}) show that square-root valued
iterated integrals and (inverse) binomial sum representations already have a close relation to elliptic 
integrals, cf.~Section~\ref{BS:sec:5}.
\subsection{Square-root valued iterated integrals: two scales}
\label{BS:sec:45}

\vspace*{1mm}
\noindent
In two-scale problems, which factorize at first order, the Feynman integrals can be represented 
in terms as iterated integrals of the kind
\begin{eqnarray}
\label{eq:GL1}
G_{l,k_1,...,k_m}(\eta;x) = \int_0^x dy~h_l(y,\eta) G_{k_1,...,k_m}(\eta;y),~~~G_\emptyset = 1,
\end{eqnarray}
where $l,k_i$ label the respective letters and $\eta$ denotes the ratio of the two scales. The 
functions $h_l(y,\eta)$ are rational or square-root valued in known 
examples~\cite{Ablinger:2017err,Ablinger:2017xml,Ablinger:2018brx}. The associated 
$N$-representations are obtained again by a Mellin transform. In some cases one finds sum 
representations, usually of generalized binomially weighted sums. However, there are also cases in 
which the $N$-space solution has no nested sum-product representation but contains higher 
transcendental functions in $N$,~\cite{Ablinger:2017xml}. A typical iterated integral of the kind 
(\ref{eq:GL1}) is given by
\begin{eqnarray}
&& G\left[\left\{\frac{\sqrt{(1-x)x}}{1-x(1-\eta)}, \frac{1}{1-x}\right\}, z\right]
=
\frac{1}{(1-\eta)^2} \Biggl[
-i \Biggl[
\eta
   \text{Li}_2\left(-\left(\sqrt{1-z}+i \sqrt{z}\right)^2\right)
\nonumber\\ &&
+\sqrt{\eta }
   \text{Li}_2\left(\frac{\left(1-\frac{i
   \sqrt{z}}{\sqrt{1-z}}\right) \sqrt{\eta
   }}{\sqrt{\eta }-1}\right)
-\sqrt{\eta }
   \text{Li}_2\left(\frac{\left(\frac{i
   \sqrt{z}}{\sqrt{1-z}}+1\right) \sqrt{\eta
   }}{\sqrt{\eta }-1}\right)
-\sqrt{\eta }
   \text{Li}_2\left(\frac{\left(1-\frac{i
   \sqrt{z}}{\sqrt{1-z}}\right) \sqrt{\eta
   }}{\sqrt{\eta }+1}\right)
\nonumber\\ &&
+\sqrt{\eta }
   \text{Li}_2\left(\frac{\left(\frac{i
   \sqrt{z}}{\sqrt{1-z}}+1\right) \sqrt{\eta
   }}{\sqrt{\eta
   }+1}\right)
+\text{Li}_2\left(-\left(\sqrt{1-z} + i \sqrt{z}\right)^2\right)\Biggr]
+(\eta -1) \sqrt{(1-z) z}
\nonumber\\ &&
+i (\eta +1) \arcsin^2\left(\sqrt{z}\right)
+(1-\eta ) \arcsin\left(\sqrt{z}\right)
+2 (\eta +1) \ln(2) \arcsin\left(\sqrt{z}\right)
+\ln(1-z) 
\nonumber\\ && \times
\Biggl[
(1-\eta ) \sqrt{(1-z) z}
+2\sqrt{\eta } \arctan\left(\frac{\sqrt{\eta z}}{\sqrt{1-z}}\right)
\Biggr]
+\ln
   \left(\frac{1-\sqrt{\eta }}{\sqrt{\eta
   }+1}\right) \Bigl(\pi  \sqrt{\eta }
-\frac{1}{2} i (\eta +1) \zeta_2
\nonumber\\ &&
-2
   \sqrt{\eta } \arctan\left(\frac{\sqrt{1-z}}{\sqrt{z}}\right)
   \Bigr)
\Biggr],
\end{eqnarray}
with the letters $\sqrt{(1-x)x}/(1-x(1-\eta))$ and $1/(1-x)$, cf.~\cite{Ablinger:2018brx}. The associated 
constants are functions of $\eta$. Furthermore, also usual harmonic polylogarithms at new algebraic arguments
contribute, cf.~\cite{Ablinger:2018brx}.
\section{Non First Order Factorizing Systems} 
\label{BS:sec:5} 

\vspace*{1mm}
\noindent
At higher and higher orders in perturbation theory, at least in the massive case, the 
system of differential equations for the master integrals will not exhibit first order factorizations
at a certain level of complexity. In various cases this will be not the case for the 
associated system of difference equations, through a Mellin transform, either. There are a series 
of well-known examples in the literature as the sun-rise integral, 
cf.~e.g.~\cite{Broadhurst:1993mw,Bloch:2013tra,Adams:2013kgc,Adams:2014vja,Adams:2015gva,Adams:2015ydq},
the kite-integral \cite{Remiddi:2016gno,Adams:2016xah}, the 3-loop QCD-corrections to the $\rho$-parameter 
\cite{Ablinger:2017bjx}, and the 3-loop QCD 
corrections to the massive operator matrix element $A_{Qg}$ \cite{Ablinger:2017ptf}, to which 
non first order factorizing systems contribute.

In decoupling the corresponding system of differential equations \cite{Zuercher:94,Ablinger:2013jta}
one obtains 2nd order factors of the associated scalar equations, which cannot be reduced further. More precisely, this associated scalar differential 
equation of order two contains more than three singularities. It is formally a Heun 
differential equation \cite{HEUN}. One can write the corresponding solution also using 
$_2F_1$-functions with rational argument \cite{IVH,Ablinger:2017bjx} and rational parameters. 
It is now interesting to see whether these solutions can be expressed in terms of complete elliptic 
integrals, which can be checked algorithmically using the triangle group \cite{TAKEUCHI}.
In the examples mentioned before one can find representations in terms of complete elliptic 
integrals of the first and second kind, {\bf K} and {\bf E}, cf.~\cite{TRICOMI,WITWAT}. A related 
question is, whether an
argument translation allows for a representation through only {\bf K}. This is possible under 
certain conditions, cf.~\cite{Herfurtner1,Movasati1}. In the case of the 3-loop QCD-corrections to
the $\rho$-parameter, however, this is not possible.

A homogeneous $_2F_1$-solution $\psi_k^{(0)}(x), k=1,2$ of the 2nd order differential equation is represented by 
the integral (\ref{eq:2F1}), which cannot be rewritten such, that it depends on $x$ just through the integral 
boundaries. This integral is therefore non-iterative w.r.t.~$x$. The inhomogeneous solution reads 
\begin{eqnarray}
\label{eq:INH}
\psi(x) = 
\psi^{(0)}_1(x)\left[C_1 - \int dx \psi_2^{(0)}(x) \frac{N(x)}{W(x)}\right]
+\psi^{(0)}_2(x)\left[C_2 + \int dx \psi_1^{(0)}(x) \frac{N(x)}{W(x)}\right],
\end{eqnarray}
with $N(x)$ and $W(x)$ the inhomogeneity  and the Wronskian and $C_{1,2}$ are the integration constants. Through
partial integration the ratio $N(x)/W(x)$ can be transformed into an iterative integral. Since  
$\psi_{k}^{(0)}(x)$ cannot be written as iterative integrals, $\psi(x)$ is
obtained as an {\it iterative non-iterative integral} 
\cite{Blumlein:2016a,Ablinger:2017bjx} of the type
\begin{eqnarray}
\label{eq:Hit}
\mathbb{H}_{a_1,...,a_{m-1};{a_m,F_m(r(y_m))},a_{m+1},...a_q}(x) &=&
\int_0^x dy_1 f_{a_1}(y_1) \int_0^{y_1} dy_2 ... \int_0^{y_{m-1}} dy_m f_{a_m}(y_m) F_m[r(y_m)] 
\nonumber\\ && \times
H_{a_{m+1},...,a_q}(y_m),
\end{eqnarray}
with $r(x)$ a rational function and $F_m$ a non-iterative integral.
Notice, that more than one non-iterative 
integral can contribute and $F_m$  denotes {\it any} non-iterative integral, implying 
a very general representation, cf.~\cite{Ablinger:2017bjx}.\footnote{This 
representation has been used in a more special form also in \cite{Remiddi:2017har} later.}
In Ref.~\cite{Adams:2018yfj} an $\varepsilon$-form for the Feynman 
diagrams of elliptic cases has been found recently. However, transcendental letters contribute here. This is in 
accordance with our earlier finding, Eq.~(\ref{eq:Hit}), which, as well is an iterative integral over all objects 
between the individual iterations and to which now also the non-iterative higher transcendental functions 
$F_m[r(y_m)]$ contribute.

One may obtain fast convergent representations of $\mathbb{H}(x)$ by overlapping series expansions around 
$x = x_0$ outside possible singularities, see Ref.~\cite{Ablinger:2017bjx} for details.

Because in many cases elliptic solutions are obtained one may transform the kinematic variable $x$ occurring as
${\rm \bf K}(k^2) = {\rm \bf K}(r(x))$ into the variable $q = \exp[i\pi \tau]$ analytically with
\begin{eqnarray}
\label{eq:EL1}
k^2 = r(x) = \frac{\vartheta_2^4(q)}{\vartheta_3^4(q)},
\end{eqnarray}
by applying a higher order Legendre-Jacobi transformation. Here $\vartheta_l, l=1,...,4$ denote Jacobi's 
$\vartheta$-functions and ${\sf Im}(\tau) > 0$. 
In this way Eqs.~(\ref{eq:INH}) and (\ref{eq:Hit})
are rewritten in terms of the new variable. The integrands are given by products of meromorphic modular 
forms, cf.~\cite{SERRE,COHST,ONO1}, which are given by linear combinations of ratios of Dedekind's 
$\eta$-function
\begin{eqnarray}
\label{eq:EL2}
\eta(\tau) = q^{\tfrac{1}{12}} \prod_{k=1}^\infty (1-q^{2k})~.
\end{eqnarray}
Depending on the largest multiplier $k \in \mathbb{N}$, $k_m$, of $\tau$ in the argument of the $\eta$-function, 
the solution transforms under the congruence subgroup $\Gamma_0(k_m)$. One can perform Fourier expansions in $q$ 
around the different cusps of the problem, cf.~\cite{ZUDILIN,BROADH18}. This representation is a uniform 
one w.r.t.\ the singularities if compared to the former $_2F_1$-representation. In the latter only three 
singularities of the problem are encoded in the $_2F_1$-function, while all others are implied by the 
argument $r(z)$ in an asymmetric way.

In the case that the occurring modular forms are holomorphic, one obtains representations in Eisenstein 
series with character, while in the meromorphic case additional $\eta$-factors in the denominators are present.
In the former case the $q$-integrands can be written in terms of elliptic polylogarithms 
in the representation
\begin{eqnarray}
\label{eq:EL3}
{\rm ELi}_{n,m}(x,y) = 
\sum_{k=1}^\infty
\sum_{l=1}^\infty \frac{x^k}{k^n} \frac{y^l}{l^m} q^{k l}
\end{eqnarray}
and products thereof, cf.~\cite{Adams:2015gva}. The corresponding $q$-integrals can be directly performed.
The solution (\ref{eq:INH}) usually appears for single master integrals. Other master integrals are obtained
integrating further other letters, so that finally representations by $\mathbb{H}(x)$ occur.
Iterated modular forms, resp. Eisenstein series, have been also discussed recently in 
\cite{Adams:2017ejb,Broedel:2018iwv}.

For systems of differential equations which factorize to 3rd or higher order much less is know. Formally, the 
representation Eq.~(\ref{eq:Hit}) also applies there. However, one will be also interested in the more specific 
properties of these systems having finally concrete cases at hand.
\section{Outlook}
\label{BS:sec:6}

\vspace*{1mm}
\noindent 
Various powerful computation methods exist to solve massless and massive problems in quantum field theory
to 4- and 5-loop order analytically for a low number of external legs and scales involved. The mathematical
understanding of the associated solution spaces made substantial progress during the last two decades and
the computer-algebraic methods gained orders of magnitude during a similar period.
One now has to see which structures have to be revealed next, going to even higher orders and/or allowing for
more scales.

Despite the fact that a systematic classification for sets of Feynman integrals can be obtained
through their degree of decoupling of the associated systems of differential equations, cf.~Section~\ref{BS:sec:2}, just 
the use of differential 
equations makes it difficult to obtain closed form analytic solutions. This is due to the fact, that 
this approach is a rather inclusive one, as also others. Already in early studies of elliptic solutions, see e.g.
\cite{Laporta:2004rb}, dispersive representations have been used. As well-known, cuts allow access to the 
specific 
integrand structures and, e.g., the elliptic solutions for the respective integrals were easily 
obtained.\footnote{For the well studied simpler one-loop case, see e.g. \cite{Abreu:2017mtm}.}
The cut-representation is equivalent to a Hilbert-transform \cite{Hilbert:1912,KRONIG,KRAMERS}
\begin{eqnarray}
F(s) = \int_{-\infty}^{+\infty} \frac{dt}{t-s} f(t)
\end{eqnarray}
splitting off one corresponding integral. In more involved cases, one might want to apply even multiple cuts to reveal 
the corresponding class  
of integrals. In this way, the dispersive approach, \cite{Veltman:1963th}, will form an important asset also 
to many key problems to be dealt with in the future. In the practical calculation of Feynman diagrams,
the general 
Hopf-algebra structure \cite{Kreimer:1997dp,Connes:1998qv,Connes:1999yr,Connes:2000fe} 
implied by the  
renormalization of the corresponding quantum field theory may play a role. In general, however,
all information is encoded in the specific multi-variate hyper-exponential structure of the individual 
integrand of the graph itself, which decides on its mathematical result and the wider class of 
mathematical structures to which 
it will finally belong. Experience shows that it is often rather difficult to determine the latter a 
priori, since these specific structures only unfold in course of the different integrations in most of 
the  interesting cases.

One can imagine that more involved Feynman diagrams could lead to Abel-integrals 
\cite{NEUMANN}. Also integrals related to K3-surfaces \cite{BSCH} are expected.
The development in the field of zero- and single-scale quantities during the last 25 years let us
expect a rich host of mathematical structures to be unrevealed and techniques to be discovered and 
designed going to even higher loop orders 
and allowing to study also cases in which a few more invariants are present during the decades to come.

The theoretical predictions for the measurements planned at the FCC have to have an accuracy
below the experimental errors which can be achieved. This will require at least one order of magnitude
better complete calculations in the respective quantum field theory, and in some important cases even 
more. To achieve this will require important improvements of what is technically possible at present, 
also in case of analytic calculations. The present experience has to be maintained for the future and 
to be significantly extended. This will also require an even closer cooperation between physicists and 
mathematicians 
and will include the solution of unprecedented tasks within computer algebra as well. As experience tells, 
these enormous tasks are by far not self-organizing and are of very long term nature to be taken into account
in the thorough planning of this collider project. Large-scale and long-term research of this kind is therefore 
often performed with a strong involvement of leading international research centers. A series of important calculations 
for HERA and the LHC spanned over {\it two decades}, which is setting a lower bound to what will be needed at the FCC. A 
substantial theoretical community  will be needed to solve these tasks. In observing the fantastic, strongly 
interdisciplinary leaps forward having been made since the late 1980ies, it seems to be possible, but a great
effort is needed to master this adventure. The theoretical and mathematical insight which will be gained will be
very large. The task is indispensable, however, to be able to interprete the experimental precision measurements 
at a facility like the FCC.

\vspace*{3mm}
\noindent
{\bf Acknowledgment.}~
We would like to thank J.~Ablinger, D.B.~Broadhurst, D.~Kreimer, and P.~Marquard for discussions.
This work was supported in part by the Austrian Science Fund (FWF) grant SFB F50 (F5009-N15). 

{\footnotesize
\bibliographystyle{utphys_spires_tit}
\bibliography{JB1}
\end{document}